\documentclass[10pt,a4paper]{article}
\usepackage[utf8]{inputenc}
\usepackage{amsmath}
\usepackage{amsfonts}
\usepackage{amssymb}
\usepackage{graphicx}
\usepackage{algorithm}
\usepackage{algpseudocode}
\usepackage{enumitem}
\usepackage{authblk}

\begin{document}

\title{Sorting sums of binary decision summands}
\author{Torsten Gross}
\author{Nils Bl\"uthgen}
\affil{Institute of Pathology, Charit\'{e} - Universit\"atsmedizin Berlin}
\affil{Integrative Research Institute for the Life Sciences and Institute for Theoretical Biology, Humboldt University of Berlin}

\date{\today}
\maketitle

\begin{abstract}
A sum where each of the $N$ summands can be independently chosen from two choices yields $2^N$ possible summation outcomes. There is an $\mathcal{O}(K^2)$-algorithm that finds the $K$ smallest/largest of these sums by evading the enumeration of all sums.
\end{abstract}

In many applications independent binary choices have to be made for each element in a set of number pairs. If the sum of the chosen numbers has to be evaluated e.g.\ as a score, it is often important to find the top scoring series of choices.
A typical scenario of such kind is the transmission of a binary sequence via a noisy channel. Then, the receiver is, at most, provided with probabilistic information about the two states of each individual bit, based on which a series of choices must be made to construct the received bit string.
This may occur for example when processing noisy voltage levels in digital electronics \cite{Shepard:1997:NDS:244522.244879}, where a measure of confidence in the on- and off-state might be quantified by the difference of the measured to the expected voltage of the according state. 
Similar problems appear in the processing of blurry images of barcodes \cite{yang2012binarization,4310076} or of binary experimental data that is subject to measurement error, among others. 
The sum over such confidences can then be interpreted as an overall confidence score, by which all possible bit strings can be ranked. 
Such ranking becomes crucial when the bit string needs to pass a validation e.g.\ by a checksum, a password or by combinatorial constraints. Then, a natural way to identify a valid message that is similar or identical to the original message is by iteratively testing candidate sequences in the order of decreasing confidence scores, such that high scores will be considered first.
Ordering binary sequences by such a sum is a sorting problem that we address in this article.\\

The problem can be formalised as follows. Let us consider $N$ pairs of real numbers. 
From each pair we are to choose one number which yields a total of $2^N$ different choice combinations. 
For each such combination we can compute a sum over its $N$ numbers. 
The goal is then to construct an algorithm which identifies the $K$ choice combinations with the $K$ smallest (or largest) sums, when $K\ll 2^N$.

A naive way to accomplish this task is to enumerate all possible combinations and sort their according sums by a suitable sorting algorithm. 
Evidently, due to the exponential growth of the number of sums with increasing $N$ this approach fails for $N\gtrapprox 25$. 
To solve the problem in polynomial time, we developed an iterative algorithm whose complexity scales quadratically in $K$ and is therefore applicable for large $N$.

\section{Decision-sums sorting algorithm}
For a given set of $N$ number pairs, we shall assemble the smaller number from each pair within vector $\mathbf{v}_0 \in \mathbb{R}^N$ and the larger one within $\mathbf{v}_1 \in \mathbb{R}^N$. From each pair we are to choose one number. The choice combination with the $i$-th smallest sum over the chosen numbers shall be denoted by a binary vector $\mathbf{c}_i \in \mathbb{B}^N$. Its zero/one components indicate, whether the smaller or the larger number was chosen for each number pair, respectively. The aforementioned sum to each choice combination is defined as 
\begin{align}\label{eq:def_S}
S(\mathbf{c}_i)= \left(\mathbf{1}-\mathbf{c}_i\right) \mathbf{v_0} +\mathbf{c}_i \, \mathbf{v}_1=\sum_{j=1}^N \left[\mathbf{v}_0 \right]_j + \mathbf{c}_i (\mathbf{v}_1 - \mathbf{v}_0)  .
\end{align}
The goal of the sorting algorithm is to identify the set of choice combinations 
\[\mathcal{C}^K=\left\lbrace \mathbf{c}_1 \ldots \mathbf{c}_K\right\rbrace \text{ with } S(\mathbf{c}_1) \leq \ldots \leq S(\mathbf{c}_K) \leq S(\mathbf{c}_l), \quad \forall\, l>K.\]
If instead we are to determine the combinations with largest sums we can do so by simply multiplying all numbers with -1 and then finding the smallest combinations. For convenience, we define 
\[
\boldsymbol{\Delta}=\hat{P}\left( \mathbf{v}_1 - \mathbf{v}_0 \right),\, \mathbf{\hat{c}}_i=\hat{P}\,\mathbf{c}_i\, \text{ and }\, \hat{\mathcal{C}}^K=\left\lbrace \hat{\mathbf{c}}_1 \ldots \hat{\mathbf{c}}_K\right\rbrace,
\]
with permutation matrix $\hat{P}$ that orders the $ (\mathbf{v}_1 - \mathbf{v}_0)$ - components by size, such that 
\begin{align}\label{eq:delta_ordering}
\Delta_i \geq \Delta_j \geq 0\text{, for }i>j.
\end{align}
When $\hat{\mathcal{C}}^K$ is determined, one can easily return to the original ordering of the components by applying the inverse permutation matrix.
 
From Eq.\,\eqref{eq:def_S} the smallest sum can readily be identified as $S(\mathbf{c}_1)\equiv S_1=\sum_j \left[\mathbf{v_0} \right]_j$ by choosing $\mathbf{\hat{c}}_1=\mathbf{c}_1=\mathbf{0}$, so that we can reformulate Eq.\,\eqref{eq:def_S} as
\begin{align}\label{eq:sum}
S(\mathbf{c}_i)=S(\hat{P}^{-1}\,\mathbf{\hat{c}_i})=S_1+\mathbf{\hat{c}_i}\,\boldsymbol{\Delta}.
\end{align}

We can now proceed to describe an iterative algorithm that finds the next choice combination in each iteration step. The algorithm makes use of shift operators $\mathbb{S}^i$ acting on a choice combination vector $\mathbf{c}=
\left[c_1\ldots c_N\right]$ as follows
\begin{align*}
\mathbb{S}^i\,\mathbf{c}=
\begin{cases}
\left[1,c_2\ldots c_N\right] &\text{if } i=1,\\
\left[\ldots c_{i-2}, c_{i}, c_{i-1},c_{i+1}\ldots \right] &\text{if } i>1, \text{ and }c_{i-1}=1,\\
\mathbf{c} &\text{else.}
\end{cases}
\end{align*}
Thus $\mathbb{S}^i$ moves one-entries from position $i-1$ to $i$ or sets the first component to one if $i=1$. Two rather apparent properties are associated with the shift operator.
\begin{enumerate}[label={P\arabic*}]
\item \textit{Any choice combination can be obtained by a chain of shift operators acting on} $\mathbf{c_1}=\mathbf{0}$.\label{enum:shift1}
\item $S(\mathbb{S}^i\,\mathbf{\hat{c}}_j)\geq S(\mathbf{\hat{c}}_j)$\label{enum:shift2}
\end{enumerate}
\textit{Proof.} To show \ref{enum:shift1} it suffices to realize that any combination vector can be constructed from the zero vector by generating a one-entry at the first position and then shifting it to its target position and repeating this procedure for each one-entry in the target vector.\\
\ref{enum:shift2} is a direct consequence of Eqs.\,\eqref{eq:delta_ordering} and \eqref{eq:sum}.\\

The idea of the algorithm is to carry out iteration steps $i=1\ldots K$ and at each step to update a small set of choice combinations $\mathcal{P}^i$, from which we can extract $\mathbf{\hat{c}}_{i+1}$. To construct $\mathcal{P}^i$ we make use of yet another set of combination vectors, $\mathcal{Q}(\mathbf{c})$, which comprises all vectors that differ from $\mathbf{c}$ after applying any single shift operator to it. That is,
\[
\mathcal{Q}(\mathbf{c})=\left\lbrace \mathbb{S}^i\mathbf{c}: i=1\ldots N  \right\rbrace \setminus \left\lbrace \mathbf{c} \right\rbrace.
\]
Note that, because $\mathbb{S}^i$ only alters $\mathbf{c}$ if $c_{i-1}=1$ and $c_i=0$,
\begin{align}
|\mathcal{Q}(\mathbf{c})|\leq N/2.\label{eq:nbr_els_in_Q}
\end{align}
Now we shall iteratively define $\mathcal{P}^i,\; i\geq 0$.
\begin{align}
\mathcal{P}^0&=\left\lbrace \mathbf{\hat{c}}_1 \right\rbrace \nonumber \\
\mathcal{P}^i&=\left(  \mathcal{P}^{i-1} \setminus \mathbf{\hat{c}}_i\right) \cup \mathcal{Q}(\mathbf{\hat{c}}_i)\label{eq:P_set_iter}\\
	&=\Big( \bigcup_{j=1}^i \mathcal{Q}(\mathbf{\hat{c}}_j) \Big) \setminus \hat{\mathcal{C}}^i\label{eq:P_set_full}
\end{align}
$\mathcal{P}^i$ has two crucial properties.
\begin{enumerate}[resume,label={P\arabic*}]
\item $\mathbf{\hat{c}}_{i+1}\in \mathcal{P}^i$\label{enum:P_set1}
\item $S(\mathbf{\hat{c}}_{i+1})\leq S(\mathbf{\hat{c}}_{j}), \; \forall \,\mathbf{\hat{c}}_{j} \in \mathcal{P}^i$\label{enum:P_set2}
\end{enumerate}
\textit{Proof.} \ref{enum:shift1} implies the existence of at least one choice combination $\mathbf{\tilde{c}}$ and a specific $j\in [1\ldots N]$, such that $\mathbb{S}^j\,\mathbf{\tilde{c}}=\mathbf{\hat{c}}_{i+1}$. From \ref{enum:shift2} we know that $S(\mathbf{\tilde{c}})\leq S(\mathbf{\hat{c}}_{i+1})$ and thus, by definition, $\mathbf{\tilde{c}} \in \hat{\mathcal{C}}^i$. Therefore, realizing that $\bigcup_{j=1}^i \mathcal{Q}(\mathbf{\hat{c}}_j)$ in Eq.\,\eqref{eq:P_set_full} comprises the choice combinations that result from applying all shift operators, $ \mathbb{S}^k$ with $k=1\ldots N$, to all elements of $\hat{\mathcal{C}}^i$ implies \ref{enum:P_set1}.\\
\ref{enum:P_set2} is evident from Eq.\,\eqref{eq:P_set_full} since all choice combinations with sums smaller than $S(\mathbf{\hat{c}}_{i+1})$ are excluded from $\mathcal{P}^i$.\\

In combination, \ref{enum:P_set1} and \ref{enum:P_set2} indicate that at the $i$-th iteration step we can determine $\mathbf{\hat{c}}_{i+1}$ as the element of $\mathcal{P}^i$ with the smallest sum. If, depending on $\mathbf{v}_0$ and $\mathbf{v}_1$, several $\mathcal{P}^i$-elements have the smallest sum, their ordering in $\hat{\mathcal{C}}^K$ is ambiguous and an arbitrary one of them is chosen as $\mathbf{\hat{c}}_{i+1}$. With this we can construct the desired algorithm.
\renewcommand{\thealgorithm}{}
\floatname{algorithm}{}

\begin{algorithm}[H]
\caption{\textbf{Decision-sums sorting algorithm}}
\begin{algorithmic}[0]
\State $\mathbf{\hat{c}},\mathbf{c}_{1} \gets \mathbf{0}$
\State $\mathcal{P} \gets \emptyset$
\For{$i\gets 1, K-1$}
	\State $\mathcal{P} \gets \mathcal{P} \cup \mathcal{Q}(\mathbf{\hat{c}})\setminus \mathbf{\hat{c}}$
	\State $\mathbf{\hat{c}} \gets \text{any single element of: } \left\lbrace \mathbf{p}_\textrm{min}: S(\mathbf{p}_\textrm{min})\leq S(\mathbf{p}),\, \forall \mathbf{p}_\textrm{min},\mathbf{p}\in \mathcal{P} \right\rbrace$
	\State $\mathbf{c}_{i+1}\gets \hat{P}^{-1}\mathbf{\hat{c}}$
\EndFor
\end{algorithmic}
\end{algorithm}

The key to the efficiency of the algorithm is that $\mathcal{P}^i$ can be constructed iteratively according to Eq.\,\eqref{eq:P_set_iter} and that $|\mathcal{P}^i|$ is small, so that finding the element with the smallest sum becomes easy. In fact, a loose upper bound for $|\mathcal{P}^i|$ can be derived from Eqs. \eqref{eq:nbr_els_in_Q} and \eqref{eq:P_set_full}
\begin{align}\label{eq:nbr_els_P^i}
|\mathcal{P}^i|\leq i(N/2-1),
\end{align}
even though typically $|\mathcal{P}^i|$ is a lot smaller, see Fig.\,\ref{fig:performance}. Additionally, it is not necessary to sort all $\mathcal{P}^i$ elements at each iteration step since the information about the order of the elements at earlier steps can be reused, as will be discussed in the next section and is depicted in Fig.\,\ref{fig:algo_overview}. Furthermore, the update $\mathcal{P}^{i-1} \rightarrow \mathcal{P}^{i}$ only requires knowledge of $\mathbf{\hat{c}}_{i}$ and no other elements of $\hat{\mathcal{C}}^i$. The algorithm is therefore memory efficient because only $\mathcal{P}^{i}$ must be stored to generate the next choice combination.

\section{Complexity}

\begin{figure}
\label{fig:algo_overview}
  \centering
  \includegraphics[width=1.\linewidth]{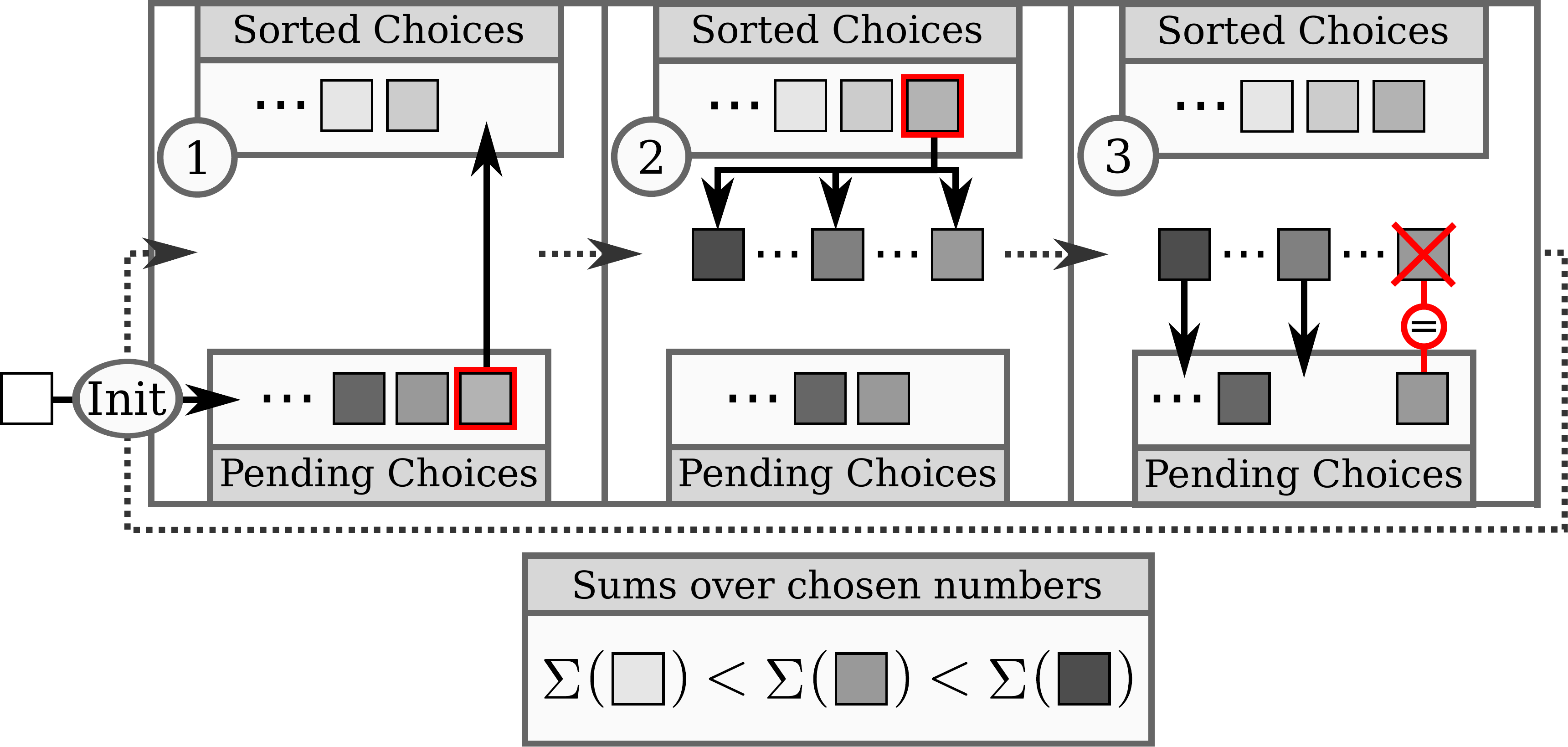}
  \caption{Implementation of the decision-sums sorting algorithm: 1. Retrieve next choice combination from pending choice combinations. 2. Generate new set of potential pending choice combinations. 3. Sort out duplicates and integrate into pending choice combinations.}
\end{figure}

The crucial advantage of the decision-sums sorting algorithm is that at each iteration step the next choice combination can be identified amongst those in the pending combination set $\mathcal{P}^i$, instead of having to identify it amongst all $\mathcal{O}(2^N)$ combinations, as is the case in the naive approach. Furthermore, it is not necessary to sort all elements of $\mathcal{P}^i$ at each iteration step $i$. If the pending combinations are stored as a sequence ordered by their according sums, only the additional elements $\mathcal{Q}(\mathbf{\hat{c}}_i)$ must be sorted and then inserted accordingly into the $\mathcal{P}^{i-1} \setminus \mathbf{\hat{c}}_i$ - sequence or duplicates be removed, as shown in Fig.\,\ref{fig:algo_overview}. Using a standard quicksort algorithm the first can be done with worst-case time complexity $\mathcal{O}(|\mathcal{Q}(\mathbf{\hat{c}}_i)|^2)$  and the latter with $\mathcal{O}(|\mathcal{P}^{i-1}|)$. The upper bounds from Eqs.\,\eqref{eq:nbr_els_in_Q} and \eqref{eq:nbr_els_P^i} therefore guarantee a complexity smaller than $\mathcal{O}(i)$ to complete the $i$-th iteration step, which yields a worst-case time complexity of $\mathcal{O}(K^2)$ for the computation of the entire sequence $\mathcal{C}^K$.

\begin{figure}
\label{fig:performance}
  \centering
  \includegraphics[width=1.\linewidth]{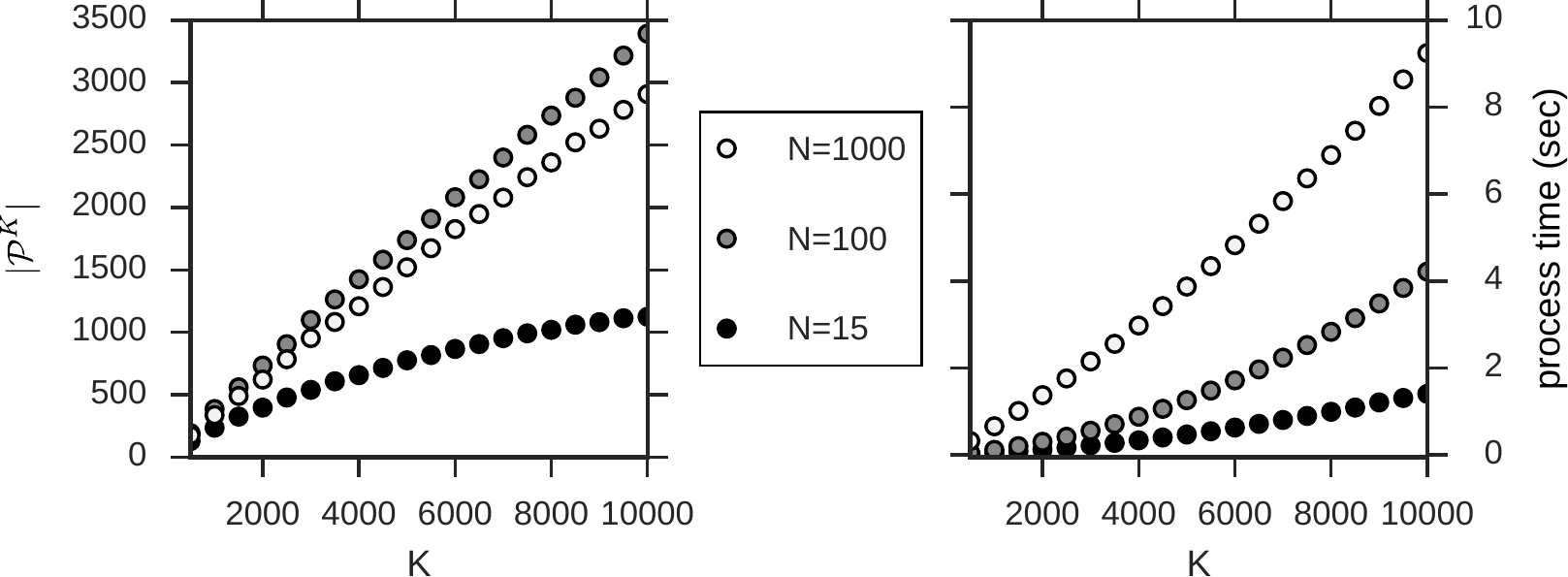}
  \caption{Performance measurements of the decision-sums sorting algorithm. Numbers in all number-pairs are sampled from the standard uniform distribution.}
\end{figure}

Let us verfify these assertions by performance measurements. We generated number-pair sets of varying size, where each number was sampled from the standard uniform distribution. Using a non-parallelized Python implementation of the decision-sums sorting algorithm, run on an Intel Core i5-6200U CPU with 2.3 GHz, we measured $|\mathcal{P}^K|$ and the process time to compute $\mathcal{C}^K$ for different $K$, see Fig.\,\ref{fig:performance}.

We observe that $|\mathcal{P}^K|$ indeed grows linearly with $K$ as long as $K/2^N\ll 1$, even though at a much lower rate than suggested by the upper bound in Eq.\,\eqref{eq:nbr_els_P^i}. When $K$ actually becomes comparable to the number of all possible choice combinations the growth of $|\mathcal{P}^K|$ can only decrease, simply because the number of potential next choices becomes more limited. This is what we observe for $N=15$. Strikingly, $|\mathcal{P}^K|$ is consistently larger for $N=100$ compared to $N=1000$ for the observed $K$ values. This behaviour was reproducible for newly sampled random numbers and thus points out that $|\mathcal{P}^K|$ grows comparatively slow for very small $K/2^N$ values.

The measured process time shows a superlinear growth with increasing $K$. For further quantification we performed least squares second degree polynomial fits on the different process time curves. For all three cases we observed $R^2>0.999$ and quadratic coefficients $9.1 \times 10^{-9}$, $3.4 \times 10^{-8}$ and $3.0 \times 10^{-8}$ for $N$ being 15, 100 and 1000, respectively. With this we see the expected quadratic scaling confirmed and note that the observed running times imply the possibility of also sorting even much larger number-pair lists.

\section{Conclusion}

We presented an algorithm to sort the sums over the combinations of numbers, where each combination selects one number from each of $N$ given number pairs. Not relying on prior computing of all $2^N$ sums, the algorithm is shown to run with a worst case complexity that is quadratic in the number of sorted combinations. However, the optimality of the algorithm was not proven so that the existence of a lower complexity bound can not be precluded. Furthermore, the decision-sums sorting algorithm could also be of theoretical interest as it relates to similar sorting problems \cite{Fredman1976355,Lambert1992137}. A Python implementation of the decision-sums sorting algorithm is freely available \cite{grosstor_algo}.



We would like to thank Dr. Manuela Benary for helpful discussions.

\bibliography{literature}{}
\bibliographystyle{plain}
\end{document}